\documentclass[prl, twocolumn]{revtex4}

\usepackage{graphicx}
\usepackage{amsmath}
\usepackage{setspace}

\def\be{\begin{equation}}
\def\ee{\end{equation}}
\def\bea{\begin{eqnarray}}
\def\eea{\end{eqnarray}}
\newcommand{\eref}[1]{(\ref{#1})}

\newcommand\refeq[1]{Eq.~(\ref{eq:#1})}

\begin{document}

\title{Calculation of Cu/Ta interface electron transmission and effect on conductivity in nanoscale interconnect technology}
\author{Baruch Feldman\footnote{Current address: Tyndall National Institute, University College Cork, Cork, Ireland}}
\affiliation{Physics Department, University of Washington, Seattle, WA}
\author{Scott T. Dunham}
\affiliation{Physics Department, University of Washington, Seattle, WA \\
Electrical Engineering Department, University of Washington, Seattle, WA}

\begin{abstract} 

Resistivity augmentation in nanoscale metal interconnects is a performance limiting factor in integrated circuits.  Here we present calculations of electron scattering and transmission at the interface between Cu interconnects and their barrier layers, in this case Ta.  We also present a semiclassical model to predict the technological impact of this scattering and find that a barrier layer can significantly decrease conductivity, consistent with previously published measurements. 

\end{abstract}

\maketitle

As the minimum feature size in transistors continues to shrink, nanoscale metal wires with thickness $\leq$32 nm will soon be needed to interconnect transistors in integrated circuits.
However, measurements show nanoscale metallic wires have substantially higher resistivity than bulk metals \cite{Munoz, GrainRefs}, leading to performance-limiting interconnect delays and power dissipation \cite{ITRS}.  
Scattering from rough wire surfaces, interfaces with liner layers, and grain boundaries are believed to be the causes of this conductivity degradation \cite{ITRS}
, but microscopic understanding of these effects and quantitative predictions of their magnitude have been limited.  In recent works, we have considered scattering from surfaces \cite{SurfScatter} and grain boundaries \cite{GrBdry}, and in this paper, we present calculations of scattering and transmission at the interface between Cu interconnects and their liner layers. 

Nanoscale Cu wires in integrated circuits are surrounded on three sides by a liner layer, typically made of Ta, TiN, or Ti.  Also known as a barrier/adhesion/seed layer, this layer acts as a seed for the deposition of Cu, improves adhesion of the Cu to the sidewalls, and prevents diffusion of Cu into the dielectric.  To inhibit the migration of Cu atoms, the liner layer is made of a refractory metal with small crystallites; this same property gives it poor conductivity despite being metallic.  In this paper, we consider for the liner layer the most common form of Ta when deposited epitaxially on Cu, $\beta-$Ta \cite{BetaTa-discover, BetaTa-epitaxy, BetaTa-resistivity}, which has a bulk resistivity of (200 $\pm$ 20) $\mu \Omega \cdot$ cm \cite{BetaTa-discover, BetaTa-resistivity, BetaTa-struct}, compared to 1.7 $\mu \Omega \cdot$ cm for Cu.  

$\beta-$Ta \cite{BetaTa-epitaxy, BetaTa-struct} has a tetragonal lattice with dimensions $a = 10.2$ \AA \; and $c = 5.3$ \! \AA.  In the $\left[ 002 \right]$ direction (corresponding to the $c$ lattice parameter), it contains four equally spaced layers, which consist of either four regularly spaced Ta atoms or 11 atoms arranged in a pseudo-hexagonal pattern.  The $\beta-$Ta unit cell contains a total of 30 atoms.  Simulations \cite{BetaTa-MD} and measurements \cite{BetaTa-epitaxy} indicate that for Cu deposited on a (002) $\beta-$Ta liner layer, the $\beta-$Ta matches heteroepitaxially to Cu (111) surfaces with a relative strain in each $a$ direction of approximately 7\%, causing the pseudo-hexagons in Ta to match those in the Cu (111) planes.  

\begin{figure}
\includegraphics[scale=0.4]{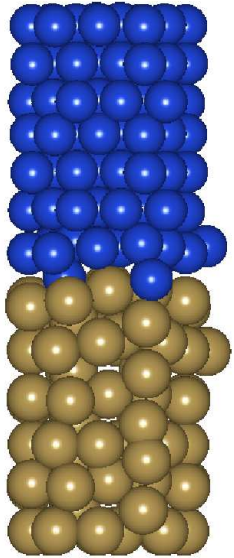}
\caption{ Relaxed Cu/Ta interface, as used in transport simulation.  Top eight monolayers (darker, blue online) are Cu, bottom (lighter, brown online) layers are Ta.  This structure simulates a Cu wire with Ta liner layer.  The wire axis is parallel to the interface (left to right, as shown).  The transport simulations use infinite Ta and Cu leads at the top and bottom of the system shown, and calculate the probability of electrons to transmit across the interface from Cu to Ta and Ta to Cu. \label{fig:epitaxy}}
\end{figure}

We have set up such an epitaxial simulation cell for relaxation with the VASP density functional theory (DFT) code \cite{VASP}.  Our interface contains a total of 56 Ta and 64 Cu atoms with periodic boundary conditions in the plane of the interface (Figure \ref{fig:epitaxy}).  These boundary conditions are made realistic by the epitaxial relationship, which should impose rough periodicity over actual interfaces.  As shown in Fig.~\ref{fig:epitaxy}, our Ta region terminates with an 11-atom Ta layer, interfacing with a partial Cu layer with four atoms, followed by four full (15-atom) Cu (111) layers.  We considered a range of interface structures, and this gave the lowest energy.   We relaxed this system within DFT, keeping the bottom Ta layer fixed and allowing the top Cu layer to adjust only in the $z$ dimension.  This allowed the interface $z-$spacing to relax.  The variation in $z-$position among Cu atoms within the topmost layer after relaxation was very small compared to the layer spacing.

To simulate transmission across a Cu/Ta interface, we replaced $z-$coordinates of the topmost Cu layer by their average, leaving ideal Cu and ideal strained $\beta-$Ta layers bounding the interface region.  We then set up a transmission simulation across the interface, using the relaxed interface of Fig.~\ref{fig:epitaxy} as a scattering region and matching the perfect Cu and Ta layers to electrodes made of ideal lattices of the respective materials.  We used the code Atomistix \cite{Brandbyge} to perform Non-Equilibrium Green's Function Method \cite{Datta} simulations of dynamical transmission across interfaces, as in our previous work \cite{GrBdry}.  

In this formulation, transmission probability 
\be
T = \bar{T}/M,
\label{eq:transmission}
\ee 
where $\bar{T}$ is the total transmission (conductance in units of $2e^2/h$), and $M$ is the Sharvin conductance in units of $2e^2/h$, which is equal to the number of current-carrying modes \cite{Datta}.  Note that conduction electron density, and therefore $M$, is different in the two materials.  We obtained $M$ for (002) strained $\beta-$Ta and (111) Cu by separate transmission simulations, finding a ratio $M_{Cu}/M_{Ta} = 1.8$.  However, reciprocity requires that $\bar{T}$ is the same for transmission in either direction across the interface \cite{Datta}, so the {\em net} current is zero across the unbiased interface after the vacuum levels have adjusted to equilibrate the Fermi levels\footnote{A space charge region must form at the interface, analogous to that at a PN-junction, to bring about this equilibrium. }.
We calculated $\bar{T}$ for transmission from Ta to Cu only, using the $M$'s to find 
an overall transmission probability $T = 0.22$ for electrons originating in Cu and 0.39 for those originating in Ta.  

Let us consider the effect on conduction of transmission across the interface.  Although the liner layer represents a parallel conductance to the Cu wire, the high resistivity of $\beta-$Ta makes this additional amount negligible \cite{Rossnagel}.  Instead, we consider the more important effect of electrons entering Ta and rapidly losing their net momentum before returning to Cu.   If electrons encounter the interface, they have probability $T$ of transmitting into Ta where they will lose their momentum.  In steady state, for every electron leaving Cu there is on average one entering from Ta.  Given the short mean free path in $\beta-$Ta, the expected value of drift velocity for returning electrons is near zero.  Effectively, electrons encounter the surface and lose their net momentum.  Thus instead of parallel conductances, a more appropriate model for the Ta barrier layer is added surface scattering in a Cu wire.  

We can estimate the effect on conduction in Cu using a simple model with the same form as the surface scattering models of Fuchs \cite{Fuchs} and Sondheimer \cite{Sondheimer}.  
Here we follow Sondheimer \cite{Sondheimer} to find the semiclassical surface encounter rate (but we interpret the momentum loss as being due to electrons returning to Cu from Ta, rather than scattering at the surface).  
We assume only a fraction $f$ of total carriers are affected by the surfaces, and as long as thickness is not too small compared to a mean free path $\lambda$, only carriers within $\sim\lambda$ of the surface should be affected. The fraction becomes:
\[
f = \frac{C P \lambda}{A},
\]
where $A$ and $P$ are the wire cross section and perimeter, and we introduce a dimensionless constant $C$.  Comparison with geometry-specific calculations gives $C = 3/16$ \cite{Sondheimer}.  The associated resistivity augmentation over bulk is given by 
\be
\label{eq:FS}
\frac{\rho}{\rho_b} = 1 + \frac{3 \: \lambda \: P}{\: 16 A \:} (1-p),
\ee
for arbitrary cross section,  
where $1-p$ is the proportion of electrons that lose their momentum on interaction with the surface.  In our case, this is the proportion of electrons transmitting into the barrier layer and losing their net drift velocity before returning to Cu. 

The assumption of zero drift velocity for returning electrons is valid when the Ta thickness is large compared with the bulk mean free path $\lambda_{\rm Ta}$ in $\beta-$Ta, which we estimate as follows.  The product of Sharvin (ballistic) conductance and bulk resistivity gives a length scale of order the bulk mean free path \cite{Datta}, a relation that holds to within 40\% for Cu\footnote{
Semiclassically, one can think of an electron traveling ballistically in between bulk scattering events.}.  
For strained (002) $\beta-$Ta, our calculation of $M_{Ta}$ combined with bulk resistivity measurements suggest $\lambda_{\rm Ta} \sim1$ nm (compared to $\lambda=39$ nm in Cu).  

We now combine \refeq{FS} with our calculated transmission probability to estimate the effect on conductivity from scattering in the barrier layer.  
Treating transmission into Ta as equivalent to a 
diffuse scattering event, the effective $(1-p)$ for a wire would be given 
by 
\be
1 - p_{\rm liner} = (1-p_s) \; \frac{P-P_i}{P} + \left\{ 1-p_i + T \right\} \frac{P_i}{P} , 
\label{eq:impact-liner-layer}
\ee
where $P$ is the total wire perimeter as in \refeq{FS}, $P_i$ is the perimeter interfacing with the liner layer, $1-p_i$ is the diffuse reflection probability from the Cu/Ta interface, and $1 - p_s = 0.04$ is our previously calculated diffuse probability for a rough Cu surface \cite{SurfScatter}.  Here we will use $p_i \approx p_s$ as a crude approximation.  
Eqs.~\eqref{eq:FS}--\eqref{eq:impact-liner-layer} give a resistivity augmentation of about 13\% for a 45 nm Cu square wire surrounded by Ta of more than about 2 nm on three sides, and 4\%  
for a 45 nm Cu film with Ta on one surface. 

We can also easily extend this analysis to consider thinner liner layers of thickness $t_{\rm Ta}$.  
The cumulative scattering probability is Poisson:
\[
P(s) = 1 - e^{-s / \lambda},
\]
where $s$ is distance traveled and $\lambda$ is mean free path.  
Then the probability that an electron leaving Ta scattered since coming from Cu is given in terms of the transmission probability from Ta, $T_{Ta} = \bar{T}/M_{Ta}$, by:

\begin{eqnarray}
1 -  e^{-2t_{\rm Ta} / \lambda_{\rm Ta}} T_{Ta}  \left[  \sum_{n=0}^\infty \left( e^{-2t_{\rm Ta} / \lambda_{\rm Ta} } (1-T_{Ta}) \right)^n  \right] \nonumber \\
 = \frac{1 - e^{-2t_{\rm Ta} / \lambda_{\rm Ta} }}{1 - (1-T_{Ta}) e^{-2t_{\rm Ta} / \lambda_{\rm Ta} }} . \nonumber
\end{eqnarray}
(Here $n$ is the number of internal reflections within Ta, which each occur with probability $1-T_{Ta}$).  
So we find
\begin{eqnarray}
1 - p_{\rm liner}(t_{\rm Ta}) \; = \; (1-p_s)  \frac{P-P_i}{P} + \nonumber \\ 
 \left\{ 1-p_i + \left( \frac{1 - e^{-2t_{\rm Ta} / \lambda_{\rm Ta} }}{1 - (1-T_{Ta}) e^{-2t_{\rm Ta} / \lambda_{\rm Ta} }} \right) T_{Cu} \right\} \frac{P_i}{P}  ,
\label{eq:impact-liner-MFP}
\end{eqnarray}
which interpolates smoothly between our surface scattering-only result and \refeq{impact-liner-layer}.  Here $T_{Cu} = \bar{T} / M_{Cu}$ is the transmission probability from Cu.  

\begin{figure}
\includegraphics[scale=0.55]{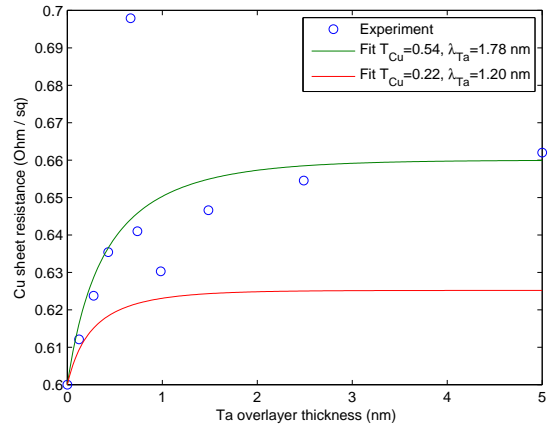}
\caption[Cu resistance {\em vs.} Ta overlayer thickness]{Increasing Cu sheet resistance as a function of Ta overlayer thickness, as measured in Ref.~\cite{Rossnagel}.   
Also shown are fits to \refeq{impact-liner-MFP},  
with $T_{Cu}$ at our calculated value of 0.22 as well as a better fit value of 0.54.  See discussion in the text regarding this quantitative discrepancy.  
\label{fig:Rossnagel}}
\end{figure}

The form of \refeq{impact-liner-MFP} agrees qualitatively with the results of Rossnagel and Kuan \cite{Rossnagel}, who measured sheet resistance of a 45 nm Cu film with 0--5 nm of Ta deposited on it (note they use a film and deposit Ta on only one surface).  They find resistivity augmentation rapidly increasing by 10\% for $\sim$2 nm of Ta, and leveling off for thicker layers.  In Figure \ref{fig:Rossnagel} we show their data, together with fits to \refeq{impact-liner-MFP} using both our calculated value of $T_{Cu} = 0.22$ and to the best-fit value of $T_{Cu} = 0.54$.  These two fits use $\lambda_{\rm Ta} \approx 1.2$ nm, and $\lambda_{\rm Ta} \approx 1.8$ nm respectively, for the mean free path in $\beta$-Ta, in agreement with our expectation of $\sim$1 nm.  
Although our calculation of $T_{Cu}$ fails to reproduce quantitatively the experimental curve in Fig.~\ref{fig:Rossnagel}, the figure shows both that the form of \eref{eq:impact-liner-MFP} agrees with experimental data, and that we can predict behavior qualitatively similar to that observed macroscopically purely from first-principles transport simulation and a semi-classical model with simple assumptions.  Moreover, Rossnagel and Kuan find that resistivity decreases back to roughly the Cu-only value upon oxidation of the interface \cite{Rossnagel}.  Thus the elimination of the Ta conducting path actually improves conduction, just as we would expect.

However, as noted, our simple analysis gives only 4\% resistivity augmentation for this geometry, compared with 10\% observed by Rossnagel and Kuan \cite{Rossnagel}.  
There are three possible reasons for the quantitative disagreement on the magnitude of resistivity augmentation.  First, there may be an underestimate in this work of the transmission probability across the interface.  Resistivity augmentation of 10\% for a single Ta overlayer would require an effective $1-p \approx 0.3$, meaning that transmission probability should be about 54\%.  Note, though, that such high $T$ is borderline inconsistent with our calculation $M_{Cu}/M_{Ta} = 1.8$.  This could be explained by an increase in the Fermi level in Ta, or by the small crystallite size in Ta, leading to many interfaces with other orientations of Ta for which $M$ is different. 

Using $p_i \approx p_s$ may well underestimate the probability of diffuse {\em reflection} at the interface.  Then
our calculation is a lower limit on scattering due to the liner layers.  
The Cu/Ta epitaxy in real samples introduces non-specular reflection from the Ta to within the angles allowed by the epitaxial periodicity, even for a perfectly periodic interface.  
Finally, it is possible that the electrons would be subject to non-specular reflection due to disorder in the interface.  
These last two explanations are also consistent with the thickness dependence in Fig.~\ref{fig:Rossnagel} because for Ta thickness less than 1 nm, the interface is not fully established.  A more detailed study of angular information on the reflected electrons at the interface would tell us the severity of non-specular reflection at the interface, and therefore clarify the last two possibilities.

In summary, we have calculated transmission probability across Cu/Ta liner layer interfaces and developed a simple model that matches qualitatively the behavior of resistivity.  This work suggests that liner layers significantly degrade the conductivity of nanoscale wires and that the use of insulating liner layers to reduce the transmission probability could significantly improve interconnect conductivity \cite{SurfScatter}. 


We would like to thank Thomas Pedersen and Dmitri Novikov for helpful discussions.  This work was supported by Intel Corporation.  


\end{document}